\documentclass[10pt,conference,a4paper]{elektron}

\usepackage[latin1]{inputenc}
\usepackage[compress]{cite}
\usepackage{graphicx}
\usepackage{amsmath}
\usepackage{amsfonts}
\usepackage{amssymb}
\usepackage{algorithmic}
\usepackage{textcomp}
\usepackage{xcolor}
\usepackage{url}

\begin{document}

\title{Using Spatial Diffusions for Optoacoustic Tomography Image Reconstruction\\ \vspace{0.5cm}
\large Uso de modelos de difusión para la reconstrucción de imágenes de tomografía optoacústica}

\author{
	\IEEEauthorblockN{Mart\'in G. Gonz\'alez\IEEEauthorrefmark{2}\IEEEauthorrefmark{1}$^1$, Mat\'ias Vera\IEEEauthorrefmark{2}\IEEEauthorrefmark{1}$^2$ and Leonardo Rey Vega\IEEEauthorrefmark{2}\IEEEauthorrefmark{1}$^3$\vspace{0.2cm}}
	
	\IEEEauthorblockA{\IEEEauthorrefmark{2}\emph{Universidad de Buenos Aires, Facultad de Ingeniería, GLOmAe}\\\emph{Paseo Colón 850, C1063ACV, Buenos Aires, Argentina}}
	
	\IEEEauthorblockA{\IEEEauthorrefmark{1}\emph{Consejo Nacional de Investigaciones Cient\'ificas y T\'ecnicas de Argentina (CONICET)}\\\emph{Godoy Cruz 2290, C1425FQB, Buenos Aires, Argentina}}
	
	\IEEEauthorblockA{$^1$\texttt{\small{mggonza@fi.uba.ar}}\\$^2$\texttt{\small{mvera@fi.uba.ar}}\\$^3$\texttt{\small{lrey@fi.uba.ar}}}
}



\maketitle

\begin{abstract}
Optoacoustic tomography image reconstruction has been a problem of interest in recent years. By exploiting the exceptional generative power of the recently proposed diffusion models we consider a scheme which is based on a conditional diffusion process. Using a simple initial image reconstruction method such as Delay and Sum, we consider a specially designed autoencoder architecture which generates a latent representation which is used as conditional information in the generative diffussion process. Numerical results show the merits of our proposal in terms of quality metrics such as PSNR and SSIM, showing that the conditional information generated in terms of the initial reconstructed image is able to bias the generative process of the diffusion model in order to enhance the image, correct artifacts and even recover some finer details that the initial reconstruction method is not able to obtain. 
\\
\\
Keywords: deep learning, photoacoustic imaging, delay and sum, image enhancement, diffusion  
\\
\\
$~~$\emph{Resumen---} La reconstrucción de imágenes por tomografía optoacústica ha sido un problema de interés en los últimos años. Aprovechando el excepcional poder generativo de los modelos de difusión propuestos recientemente, consideramos un esquema que se basa en un proceso de difusión condicional. Utilizando un método de reconstrucción de imagen inicial simple como \emph{Delay and Sum}, consideramos una arquitectura de autocodificador especialmente diseñada que genera una representación latente que se utiliza como información condicional en el proceso de difusión generativo. Los resultados numéricos muestran los méritos de nuestra propuesta en términos de métricas de calidad como PSNR y SSIM, mostrando que la información condicional generada en términos de la imagen reconstruida inicial es capaz de sesgar el proceso generativo del modelo de difusión para mejorar la imagen, corregir artefactos e incluso recuperar algunos detalles más finos que el método de reconstrucción inicial no es capaz de obtener.
\\
\\
Palabras clave: aprendizaje profundo; imágenes fotoacústicas; retardo y suma, mejora de imagen, disusión
\end{abstract}


\IEEEpeerreviewmaketitle

\section{Introduction}
\label{sec:intro}
Optoacoustic tomography (OAT) imaging uses laser excitation and ultrasonic sensors in order to provide high contrast images of biological tissues\cite{Kruger_Liu_Fang_Appledorn_1995, xu2006, rosenthal2013}. As the laser illuminates a biological tissue, the heat absorbed produces pressure waves that are captured by ultrasonic detectors, which are placed on a ring around the sample \cite{Minghua_Xu_Wang_2002, paltauf2017}. The set of all signals captured by sensors (the \emph{sinogram}) are sampled by a high-sampling rate analog-digital converter and fed into a computed, where appropriate numerical algorithms try to reconstruct the original pressure distribution due to the laser light absorption by the illuminated sample. This is known as the \emph{acoustic inversion problem} \cite{rosenthal2013, Hauptmann_Cox_2020}. After this initial distribution pressure problem is obtained, the so-called \emph{optical inversion problem} is solved in order to obtain a map of the optical absorption in the tissue. As optical absorption is usually linked with important properties, among others oxygen saturation and hemoglobin concentration, that have important diagnostic value \cite{Laufer_Delpy_Elwell_Beard_2006, Hauptmann_Cox_2020}, the optical reconstruction problem is very important in practice. Despite this, in this paper we will concentrate our efforts in the acoustic inverse problem as it is the first stage in the image recovery problem, and because from the initial pressure image a lot of information of the sample to be studied is recovered. 

Several well-known inversion procedures are \emph{model-based} ones, i.e., they start from a mathematical model that represents the forward transformation of the initial pressure distribution into the sinogram. From that model, the inverse algorithm is derived and applied to the sinogram for image retrieval for the initial pressure distribution. This approach requires a precise modeling of the experimental setup used for capturing the sinograms. It is well-known that in general, the forward model is ill-conditioned and the obtained images can present artifacts, noise amplifications, etc \cite{Hauptmann_Cox_2020}. This is aggravated by the fact that uncertainties in the forward model are common, e.g: sensor positions, speed of sound in the sample, etc \cite{hirsch2021}.

In recent years, deep learning techniques \cite{LeCun_Bengio_Hinton_2015} gained widespread attention. These are \emph{data-driven} structures, where a deep neural network is optimized using measured and/or synthetic input-output pairs. Although the optimization is computationally demanding, the results are highly competitive with those of standard approaches. Moreover, the neural architectures, if properly designed, are robust to uncertainties in the forward model \cite{vera2023}. Recently, some variational deep learning techniques, the so-called Diffusion Probabilistic Models \cite{Ho_Jain_Abbeel_2020} have received significant attention for their excellent properties in generative tasks. Inspired by non-equilibrium thermodynamics \cite{Sohl-Dickstein_Weiss_Maheswaranathan_Ganguli_2015}, these architectures sample high-quality and high-dimensional images by learning to reverse a forward diffusion process that contaminate the original images by the aggregation of controlled noise. In this way, after the model is trained, high-quality images of the same type as the ones used to adjust the weights of the structure, can be sampled with great quality and with controlled computational complexity. This structures can also be easily adapted to construct conditional models, where the reverse process can also use additional information which is linked with the desired generated images, like image labels information, text prompts, or a low resolution image for super-resolution image generation \cite{ho2022,Rombach_Blattmann_Lorenz_Esser_Ommer_2022}. These conditional models are also suitable for solving inverse problems as the one we are interested in.

In the present work, we will explore the use of a diffusion architecture to improve image quality reconstruction of a first initial reconstruction by a sufficiently simple to implement method. Our proposal will include an initial stage in which the sinogram is converted into a first reconstructed image. Then, this reconstruction is used as conditional information in a generative diffusion model to improve the quality of the image, recovering details not visible in the initial reconstruction. In Section \ref{sec:reconstruction} we summarize the different mathematical materials and methods used in our proposal. In Section \ref{sec:proposal} we describe our proposed method. In Section \ref{sec:numerical} numerical results for our proposal are presented. Finally, in Section \ref{sec:conclu}, some concluding remarks are discussed.

\section{Material and methods}

\subsection{Image reconstruction in TOA}
\label{sec:reconstruction}
When a pulse laser $\delta(t)$ illuminates the sample, the acoustic pressure $p(\mathbf{r},t)$ at position $\mathbf{r} \in\mathbb{R}^3$ and time $t$, satisfies the following wave equation:

\begin{equation}
\left(\frac{\partial^2}{\partial t^2} - v_s^2 \, \nabla^2 \right) p(\mathbf{r},t) = 0
\label{eq:waveeq}
\end{equation}

\noindent with the initial conditions,

\begin{equation}
p(\mathbf{r},0) = p_0(\mathbf{r})\, \text{,} \quad \left(\partial p /\partial t\right)(\mathbf{r},0)= 0 
\label{eq:waveeq_cond_ini}
\end{equation}

\noindent where $p_0(\mathbf{r})$ is the initial OA pressure and $v_s$ represents the speed of sound in the medium, which is assumed acoustically non-absorbing and homogeneous. Under the usual hypothesis of thermal and acoustic confinement \cite{Kruger_Liu_Fang_Appledorn_1995}, that is, when the laser pulse duration is short enough such that the heat conduction and acoustic propagation into neighboring regions of the illuminated region can be neglected, the initially induced pressure $p_0(\mathbf{r})$ is proportional to the total absorbed optical energy density. 

\begin{figure}
	\centering
	\includegraphics[width=8.5cm]{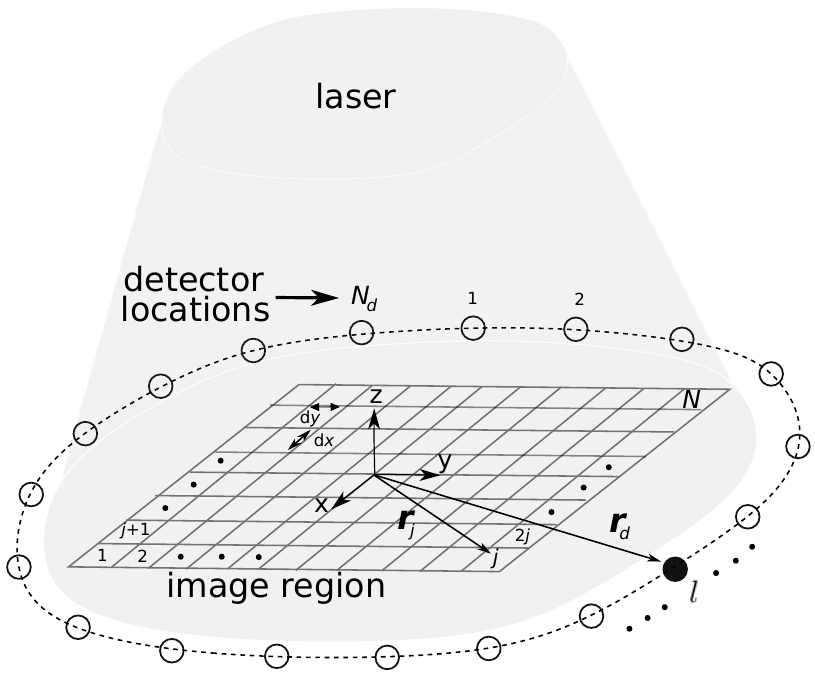}
	\caption{Schematic of the OAT imaging setup studied in this work. A number of $N_d$ detector locations are uniformly distributed around the sample, which is divided in an imaging grid of $N$ pixels.}
	\label{fig:setup}
\end{figure}
The acoustic inversion problem assumes that the wave pressure signal is measured in $N_d$ detectors as shown in Fig. \ref{fig:setup}, and those signals are used to reconstruct $p_0(\mathbf{r})$. This problem has been extensively studied in the literature. Most of the existing methods identify that the problem to be solved is an inverse one, in which we need to invert a forward operator that converts the initial pressure distribution into a sinogram \cite{rosenthal2013}. Among the most simple methods, the so-called \emph{delay and sum} (DAS) is a very easily implementable one \cite{Spadin_Jaeger_Nuster_Subochev_Frenz_2019, Ma_Peng_Yuan_Cheng_Xu_Wang_Carson_2020}. This method is a time-domain method that does not require any modeling assumption besides the knowledge of the speed of sound of the sample and the locations of the detectors. Assuming that we are interested in reconstructing a pixel at position $\mathbf{r}$ for the initial pressure distribution, we can write:
\begin{equation}
\label{eq:DAS}
S(\mathbf{r})=\sum_{d=1}^{N_d} s(d,\tau_d(\mathbf{r})),
\end{equation}
where $s(d,t)$ is the detected signal at detector $d$ and time $t$ and:
\begin{equation}
\label{eq:tau}
\tau_d(x,y)=\frac{\|\mathbf{r}-\mathbf{r}_d\|}{v_s},
\end{equation}
where $\mathbf{r}_d$ is the position of detector $d$. In colloquial terms, the method indicates that the signals of all detectors along the detection curve has to be \emph{delayed} by a quantity that depends on the distance of each detector to the corresponding image pixel and the assumed speed of sound in the sample. Then, the signals are summed up. The amplitude of the resulting signal $S(\mathbf{r})$ is representative of the strength of the initial pressure distribution $p_0(\mathbf{r})$. Although an SNR improvement of order $N_d$ is ideally achieved with this method, the truth is that several shortcomings are also present. For example, strong interference from pixels that are at the same distance to a given detector, induces reconstruction defects, artifacts and spurious details, also reducing resolution.

\subsection{Diffusions}
\label{sec:diffusions}
Probabilistic diffusions are basically latent variable models that can be factorized as a Markov chain \cite{Ho_Jain_Abbeel_2020}:
\begin{equation}
p_{\theta}(\mathbf{x}_0,\dots,\mathbf{x}_T)=p_{\theta}(\mathbf{x}_T)\prod_{t=1}^Tp_\theta(\mathbf{x}_{t-1}|\mathbf{x}_t),
\label{eq:reverse_model}
\end{equation}

\noindent where the original data $\mathbf{x}_0$ is distributed according to some unknown distribution $q(\mathbf{x}_0)$, and $\mathbf{x}_1,\dots,\mathbf{x}_T$ are intermediate latent representations, which are noisy versions (increasing noise with $t$) of $\mathbf{x}_0$. The conditional probability distributions $p_\theta(\mathbf{x}_{t-1}|\mathbf{x}_t)$ are usually modeled as multivariate Gaussian distributions $p_\theta(\mathbf{x}_{t-1}|\mathbf{x}_t)=\mathcal{N}\left(\boldsymbol\mu_t(\mathbf{x}_t,\theta),\boldsymbol\Sigma_t(\mathbf{x}_t,\theta)\right)$, where the mean and covariance matrix $\boldsymbol\mu_t(\mathbf{x}_t,\theta)$ and $\boldsymbol\Sigma_t(\mathbf{x}_t,\theta)$ are represented by deep neural networks with trainable parameters denoted by $\theta$. The distribution $p_{\theta}(\mathbf{x}_T)$ is usually kept fixed without trainable parameters and chosen as $\mathcal{N}\left(\mathbf{0},\mathbf{I}\right)$, where $\mathbf{I}$ is the identity matrix of appropriate dimensions. The full distribution $p_{\theta}(\mathbf{x}_0,\dots,\mathbf{x}_T)$ is known as the \emph{reverse process} and the marginal distribution $p_\theta(\mathbf{x}_0)$ is the ultimate-desired object, as it should be close to the true distribution of the data.  As models $p_\theta(\mathbf{x}_{t-1}|\mathbf{x}_t)$ are Gaussian, starting from  $\mathbf{x}_T\sim\mathcal{N}\left(\mathbf{0},\mathbf{I}\right)$ we can easily sample, according to $p_\theta(\mathbf{x}_{t-1}|\mathbf{x}_t)$, for $t=T,\dots,1$ and at the end generate a sample $\mathbf{x}_0$ with the same distribution of the observed data. 
In order to accomplish this, we can formulate a \emph{forward model} as:
\begin{equation}
q(\mathbf{x}_1,\dots,\mathbf{x}_T|\mathbf{x}_0)=\prod_{t=1}^T q(\mathbf{x}_t|\mathbf{x}_{t-1})
\label{eq:forward_model}
\end{equation}

\noindent where $q(\mathbf{x}_t|\mathbf{x}_{t-1})$ has no trainable parameters and modeled as $\mathcal{N}(\sqrt{1-\beta_t}\mathbf{x}_{t-1},\beta_t\mathbf{I})$, with $\beta_t$ is a predefined sequence of real numbers collectively known as $\emph{noise schedule}$ \cite{Nichol_Dhariwal_2021}. The forward model starts from a known sample $\mathbf{x}_0$ with the unknown distribution $q$ and generate a sequence of noisy versions (the amount of noise controlled by $\beta_t$) $\mathbf{x}_t$ until the step $T$, where any information of the sample $\mathbf{x}_0$ is totally destroyed, i.e., $\mathbf{x}_T\sim\mathcal{N}\left(\mathbf{0},\mathbf{I}\right)$. It is this final sample that will be the input of the reverse model.  With this forward model, we can set up the following loss function to obtain the parameters $\theta$ of the reverse model (see the mathematical details in \cite{Ho_Jain_Abbeel_2020}):

\begin{align}
L(\theta)=\frac{1}{N}\sum_{n=1}^N ((\sum_{t>1}^T D_{\rm KL}(q(\mathbf{x}_{t-1}|\mathbf{x}_t,\mathbf{x}_0^{n})\,||\,p_{\theta}(\mathbf{x}_{t-1}|\mathbf{x}_t))\nonumber\\
-\log p_{\theta}(\mathbf{x}_0^n|\mathbf{x}_1))
\label{eq:cost_function}
\end{align}

\noindent where $\left\{\mathbf{x}_0^n\right\}_{n=1}^N$ is the training set (distributed according to the unknown distribution $q(\mathbf{x}_0)$), $D_{KL}(\cdot)$ is the usual Kullback-Leibler distance between probability distributions \cite{Cover_Thomas_2006}, and $q(\mathbf{x}_{t-1}|\mathbf{x}_t,\mathbf{x}_0^{n})$ can be easily obtained from (\ref{eq:forward_model}) and shown to be a Gaussian multivariable with mean depending on $\mathbf{x}_t$, $\mathbf{x}_0^n$ and $\beta_t$ and diagonal covariance matrix depending on $\beta_t$. As all distributions are Gaussian, Kullback-Leibler distances can be computed in closed form, $\log p_{\theta}(\mathbf{x}_0|\mathbf{x}_1)$ is also easily calculated and (\ref{eq:cost_function}) becomes a quadratic cost function in terms of the training set that can be optimized for $\theta$ using standard backpropagation and gradient descent techniques. The loss function is finally written as (assuming that $\boldsymbol\Sigma_t(\mathbf{x}_t,\theta)=\sigma_t^2\mathbf{I}$ with $\sigma_t$ fixed):

\begin{equation}
L(\theta)=\frac{1}{N}\sum_{n=1}^N \sum_{t=1}^T \gamma_t\|\boldsymbol{\epsilon}_t-\boldsymbol{\epsilon}_\theta(\overbrace{\sqrt{\bar{\alpha_t}}\mathbf{x}_0^n+\sqrt{1-\bar{\alpha}_t}\boldsymbol{\epsilon}_t}^{\mathbf{x}_t},t)\|^2 ,
\label{eq:cost_function_final}
\end{equation}

\noindent where $\gamma_t$ is a parameter that depends on sequences $\beta_t$ and $\sigma_t$, $\bar{\alpha_t}=\prod_{s=1}^{t}(1-\beta_s)$, $\boldsymbol{\epsilon}_t$ is the Gaussian corrupting noise sampled from $q(\mathbf{x}_t|\mathbf{x}_{t-1})$ in the forward $t$-step, and $\boldsymbol{\epsilon}_\theta(\sqrt{\bar{\alpha_t}}\mathbf{x}_0^n+\sqrt{1-\bar{\alpha}_t}\boldsymbol{\epsilon}_t,t)$, which can be obtained from $\boldsymbol\mu_t(\mathbf{x}_t,\theta)$,
is the \emph{denoising noise} in the reverse $t$-step introduced by $p_{\theta}(\mathbf{x}_{t-1}|\mathbf{x}_t)$.

As can be seen in \eqref{eq:cost_function_final}, in each reverse $t$-step a neural network $\boldsymbol{\epsilon}_\theta$ with input $\sqrt{\bar{\alpha_t}}\mathbf{x}_0^n+\sqrt{1-\bar{\alpha}_t}\boldsymbol{\epsilon}_t$ tries to estimate the noise $\boldsymbol{\epsilon}_t$ generated in the forward $t$-step.

This is the basic idea behind a diffusion-based generative model. However, in inverse problems (which is our case) we are interested in the reconstruction of $\mathbf{x_0}$ based on the measurement of sample $\mathbf{y}_0$ which are statistically related through an unknown conditional distribution $q(\mathbf{x}_0|\mathbf{y}_0)$. For example, for our application, $\mathbf{x}_0$ would represent the true image and $\mathbf{y}_0$ the measured sinogram or some initial processing of it. The idea behind using $\mathbf{y}_0$ is that, by exploiting the statistical dependence with the desired object $\mathbf{x}_0$, the diffusion process can be efficiently ``biased'' to better capture the statistical information shared by them. In order to cover this situation, we can easily modify the above criterion and consider a reverse conditional model \cite{ho2022, Rombach_Blattmann_Lorenz_Esser_Ommer_2022} $p_{\theta}(\mathbf{x}_{t-1}|\mathbf{x}_t,\mathbf{y}_0)$ for $t=1,\dots, T$, that is also modeled by a multivariable Gaussian with trainable mean and covariance $\boldsymbol\mu_t(\mathbf{x}_t,\mathbf{y}_0,\theta)$ and $\boldsymbol\Sigma_t(\mathbf{x}_t, \mathbf{y}_0,\theta)$. The loss function can be finally written as:
\begin{align}
L(\theta)=\frac{1}{N}\sum_{n=1}^N \sum_{t=1}^T \gamma_t\|\boldsymbol{\epsilon}_t-\boldsymbol{\epsilon}_\theta(\sqrt{\bar{\alpha_t}}\mathbf{x}_0^n+\nonumber\\
+\sqrt{1-\bar{\alpha}_t}\boldsymbol{\epsilon}_t,g_\theta(\mathbf{y}_0^n),t)\|^2 ,
\label{eq:cost_function_final2}
\end{align}

\noindent where $\left\{\mathbf{x}_0^n,\mathbf{y}_0^n\right\}_{n=1}^N$ is the training set and $g_\theta$ is a trainable representation of the conditional information $\mathbf{y}_0$ for efficient processing in the reverse diffusion process. 
\begin{figure*}
	\centering
	\includegraphics[width=15cm]{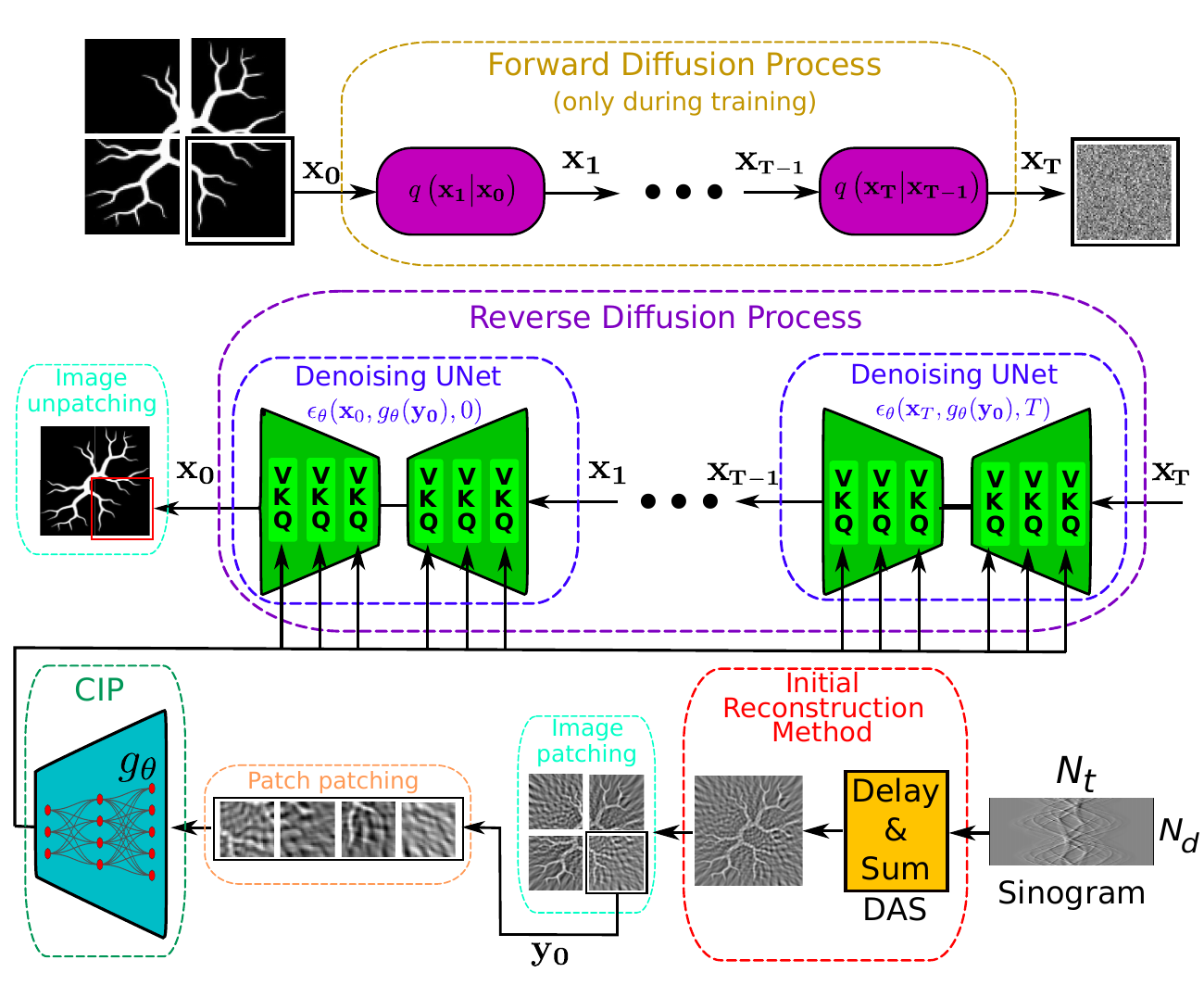}
	\caption{Proposed method architecture. The forward diffusion process only acts during  the training phase. $3-$tuples $(\mathbf{V},\mathbf{K},\mathbf{Q})$ indicate the multi-head cross-attention mechanisms at each U-Net's scales. Sinusoidal positional time embeddings are used for representing the time-steps in the reverse diffusion process. There are two levels of image patching: images delivered by DAS ($128\times 128$) patched in four smaller images ($64\times 64$). Each of these is further patched in 4 subpatches ($32\times 32$) which are processed sequentially by CIP.}
	\label{fig:arch}
\end{figure*}

\section{Proposed method}
\label{sec:proposal}
We propose to use a conditional diffusion model to enhance the image reconstructed with a prior method, which we seek to be a well-established one and easy to implement. We are not interested in a method that achieves state-of-the-art performance, but only a simple proxy between the measured sinogram and an initial image of the initial pressure distribution. The diffusion model will be in charge of significantly improving this initial estimate, being able to eliminate artifacts, increasing resolution of finer details, etc. Three major blocks can be identified in the proposed method: 1) the initial well-proved reconstruction method, 2) the conditional information preprocessing (CIP), and 3) the conditional diffusion model in reduced dimension. In Fig. \ref{fig:arch} we can see a depiction of the different blocks and their interconnections.

\subsection{Initial reconstruction method}

We choose the DAS as the initial reconstruction method. Although other possibilities can be explored \cite{Gonzalez_Vera_Dreszman_Vega_2024}, we chose it because it is simple to implement, well-studied in the literature, and because we expect that the diffusion model solves its quality problems (see Section \ref{sec:reconstruction}). 

\subsection{Conditional information preprocessing (CIP)}
\label{sec:CIP}

The initial reconstructed image from the previous block will be used as conditional information for the diffusion process. This information is given to the diffusion model using a function $g_\theta$ where $\theta$ is a set of parameters to be optimized. In our case, we considered $g_\theta$ to be the encoder part from an autoencoder with 3 hidden layers with sizes of 768, 512 and 256 and ReLu activations, respectively. The autoencoder was trained in such a way that the decoder can reconstruct the input minimizing the mean square error (MSE) loss. The output of the encoder serves as conditional information for the diffusion model and the decoder is discarded. As the diffusion model could be hard to implement in original image space, because of the high memory requirements, we considered that patching the images delivered by DAS could be useful. In this sense, the original image delivered by DAS (in our case of $128 \times 128$) is patched in four smaller images (each with size of $64\times 64$) which are further patched by a factor of four (each with size of $32\times 32$) and processed by the CIP in sequential way generating a latent representation of each patch with a size $1024 \times 1$. This latent representation for each patch is used by the diffusion model as conditional information. This same process is done not only by the reverse diffusion model but also by the forward one as shown in Fig. \ref{fig:arch}. In this last case, only the first level of patching (patches of size $64\times 64$) occurs over the ground-truth image.

\subsection{Conditional diffusion model in patch space}
\label{sec:conditional}

The conditional diffusion model generates refined reconstructions for each image patch using the conditional information generated as explained above from initial image reconstruction. During the training stage, samples corresponding to the patches of the ground-truth images are degraded during $T$ steps by the forward diffusion process, which has no-trainable parameters and where the main design choice is the noise schedule $\left\{\beta_t\right\}_{t=1}^T$. We considered a linear sequence starting from $\beta_1$ which is typically a small value (e.g. $10^{-4 }$) and ending in a larger $\beta_T$ (e.g. 0.02). The choice of $T$ proves to be critical, as large $T$ will provide a more controlled noise corruption during forward process. However, during inference the same number of $T$ sequential reverse step is to be implemented to obtain each of the final reconstructed image patches. This cannot be implemented in parallel, leading to large processing times for inference. To avoid these limitations we considered the approach taken by \cite{Song_Meng_Ermon_2022}, where some clever tricks for the sampling procedure of $p_{\theta}(\mathbf{x}_{t-1}|\mathbf{x}_t,g_\theta(\mathbf{y}_0))$ for $t=1,\dots, T$ during inference are proposed and implemented, carefully designing a fully equivalent non-Markovian processing for the forward process, that lead to Markovian reverse process with a smaller number of inference steps (NIS), speeding up inference task. Further details can be found in \cite{Song_Meng_Ermon_2022}\footnote{See also \url{https://huggingface.co/docs/diffusers/api/schedulers/ddim} for implementation details.}.

\begin{figure*}
	\centering
	\includegraphics[width=17cm]{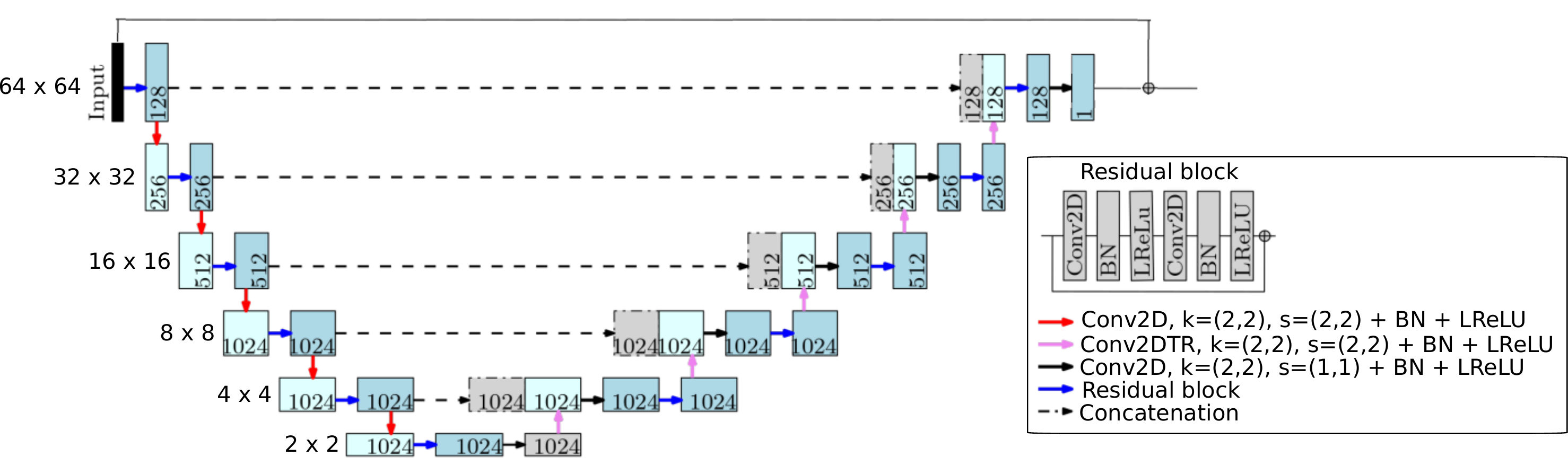}
	\caption{U-Net architecture used as benchmark to compare with our proposal.}
	\label{fig:UNet}
\end{figure*}

\begin{table}[t]
	\centering \caption{Performance (mean value and standard deviation) over the testing set.}
	\begin{tabular}{ccc}
		\hline
		Method & SSIM & PSNR \\
		\hline
		Ours (NIS 50) & \textbf{0.886} $\pm$ 0.095 & \textbf{24.045} $\pm$ 7.384 \\
		DAS & 0.185 $\pm$ 0.042 & 9.337 $\pm$ 0.738 \\
		DAS+U-Net  & 0.459 $\pm$ 0.102 & 22.884 $\pm$ 2.220 \\
		\hline
	\end{tabular}
	\label{table:1}
\end{table}

\begin{figure*}
	\centering
	\includegraphics[width=17cm]{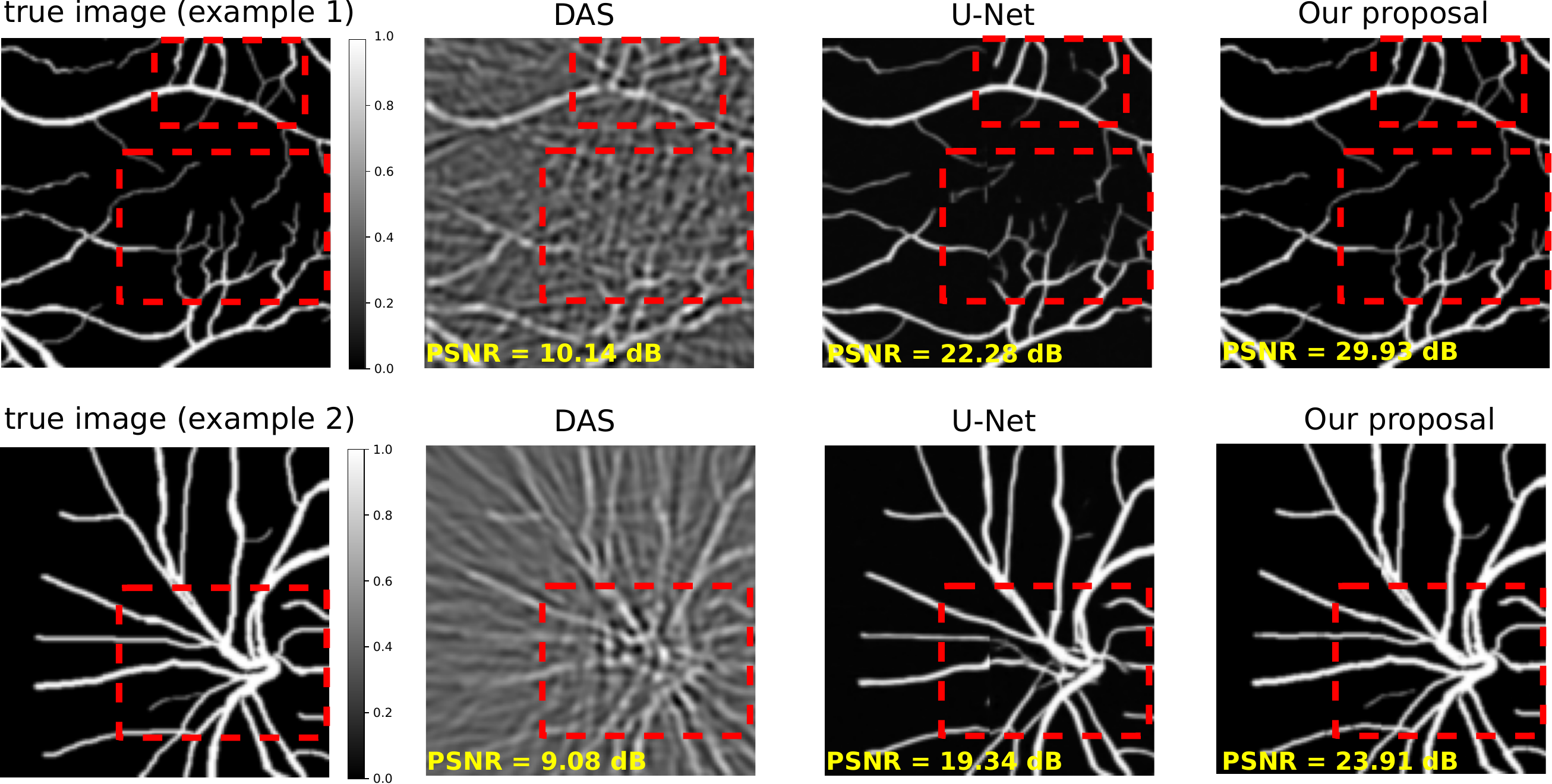}
	\caption{Reconstruction examples for our proposal (NIS = 50), DAS and UNet and their respective PSNR. Red boxes indicate finer details captured by our proposal with respect to the U-Net used as benchmark. The sinogram SNR values for examples 1 and 2 were set to $50 \text{ dB}$ and $30 \text{ dB}$, respectively.}
	\label{fig:simulresults}
\end{figure*}


\begin{figure*}
	\centering
	\includegraphics[width=17cm]{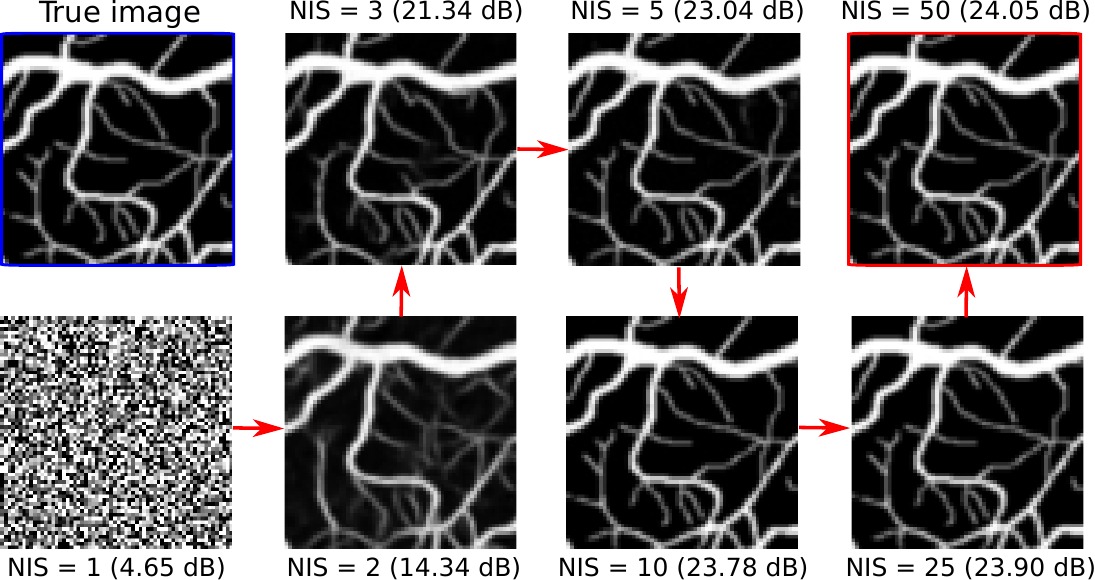}
	\caption{Sequence of images generated by our proposal at some inference time steps. Starting from NIS=1 and up to NIS=50. We also show the respective PSNR for each of the reconstructed images.}
	\label{fig:nisevol2}
\end{figure*}

The conditional information corresponding to the initial image reconstruction patches is introduced in the reverse diffusion process as explained above (i.e. $p_{\theta}(\mathbf{x}_{t-1}|\mathbf{x}_t,g_\theta(\mathbf{y}_0))$ ). The cost function given by (\ref{eq:cost_function_final2}) is then optimized using backpropagation techniques. Each step $t=1,\dots, T$ in the reverse process (i.e. $\epsilon_\theta(\mathbf{x}_t,g_\theta(\mathbf{y_0}),t)$) is modeled by a UNet architecture.
U-Net parameters are being shared across  different time-steps $t=1,\dots, T$ through a sinusoidal time embedding, as proposed in \cite{Ho_Jain_Abbeel_2020}. At each scale, a number of ResNet layers  are also used to improve training stability. The different scales of the U-Net also includes the conditional information $g_\theta(\mathbf{y_0})$ (both at the encoder and the decoder) through the use of a multi-head cross-attention structure \cite{Vaswani_Shazeer_Parmar_Uszkoreit_Jones_Gomez_Kaiser_Polosukhin_2017}. After all, patches for a given image are processed by the reverse diffusion model and then the final image is assembled as shown in Fig. \ref{fig:arch}.

\section{Results}
\label{sec:numerical}

\subsection{Implementation details}
\label{subsec:details}

To test the proposed method, we used a setting similar to the one in Fig. \ref{fig:setup} with images of $128\times 128$ pixels (pixel size of $115\, \mu\text{m}$) and $N_d=36$ detector locations uniformly distributed around the sample. The distance from the center of the sample to the circle where the sensor is positioned ($\mathbf{r}_{d}$) is $44 \text{ mm}$. The sampling frequency was set to $41 \text{ MHz}$. The detector locations has a random uncertainty of position of $0.1 \%$.  Each pressure signal has a duration of $25 \,\mu\text{s}$ ($N_t=1024$ samples). In order to generate the sinograms corresponding to the ground-truth images needed to train the proposed architecture, we made use of the program j-Wave\footnote{\url{https://ucl-bug.github.io/jwave/index.html}} \cite{Stanziola_Arridge_Cox_Treeby_2023}, which is a highly-customizable acoustics simulator based on Python and JAX.

The generated sinograms were contaminated with white Gaussian noise with different variance levels, leading to sinogram measurement SNR values between $35 \text{ dB}$ and $75 \text{ dB}$. We also considered, for DAS, a speed of sound of $v_s = 1490 \text{ m/s}$.

As explained in the previous section, each image (from DAS and ground-truth) was patched in 4 non-overlapping patches of $64 \times 64$. As represented in Fig. \ref{fig:arch}, each patch is further subdivided in four patches of $32\times 32$, each of which are flattened to vectors of size 1024 and processed sequentially by CIP, which deliver a vector of size 256 for each subpatch. Concatenation of the four subpatches gives rise to a vector of size 1024 which is used as conditional information in the reverse diffusion model, which works at a patch (of size $64\times 64$) level.   This was done to reduce the size of the diffusion model and its computational requirements for training. 

The number of $T$ chosen as $1000$ and the noise schedule start at $\beta_1=10^{-4}$ and linearly increase to $\beta_T=0.02$. For training, we used public datasets of retinal vasculature phantoms \cite{drive2020, aria2006, rite2013, stare2000, hatamizadeh2022ravir}. The number of phantoms used for training were $64,000$ and $10,000$ were used for validation.  We used ADAM optimization with $\alpha_1=0.9$ and $\alpha_2=0.999$. The initial learning rate, number of epochs and the batch size were set to $10^{-4}$, 200 and 16, respectively. The total training time was approximately 7 days in a computer with CPU Intel i9-10900F, 128 GB of RAM and two GPU RTX  3090 each with 24 GB of memory.

\subsection{Numerical results}
\label{subsec:numerical}
Besides our proposal, we also considered the following reconstruction schemes. In the first place, we considered a standalone DAS. This baseline is important because it allows us to easily see the improvements delivered by CIP + Diffusion model in the final image reconstruction. In second place, we considered the U-Net model \cite{Ronneberger_Fischer_Brox_2015}, which takes as input the initial image reconstructed by DAS. U-Net architecture has proven his worth in different tasks in image processing, including image reconstruction and enhancement \cite{vera2023}. In Fig. \ref{fig:UNet} the implemented U-Net architecture details are presented.

In table \ref{table:1} we see the average results of our proposal, DAS and DAS+U-Net when using a testing sample of $600$ images not seen during training. For the implementation of our proposal at inference we considered a NIS equal to 50, which delivers a nice trade-off between quality of the final reconstruction and computational complexity at testing. We used the following metrics: the Structural Similarity Index (SSIM) and the Peak Signal to Noise Ratio (PSNR), which are popular metrics to evaluate the performance of image reconstruction algorithms. While a SSIM is very popular perceptual metric, PSNR is a metric which objectively quantifies the level of distortion in the reconstructed image \cite{Blau_Michaeli_2018}. We see that the improvement of our proposal in both metrics with respect to DAS are significant in both metrics. For the case of DAS+U-Net we see that the corresponding metric improves significantly with respect to DAS. However, our proposal still performs better. 

In order to test our proposal in a more qualitative manner we present Fig. \ref{fig:simulresults}. We see that, although raw details are captured by DAS, there exists multiple defects and artifacts. For DAS+UNet the quality of the images are significantly improved, although some finer details (see red boxes) in the ground-truth images are not clearly recovered. Our proposal presents the best quality being able to recover almost all finer details present in the ground-truth images. 

In Fig. \ref{fig:nisevol2} it can be appreciated the evolution of the reconstructed images by our proposal when the number of NIS is sequentially increased. When NIS=1, the reconstructed image is random white noise as expected. With increasing NIS, our proposal steadily provides better reconstructed images. For NIS as small as 2, almost all structures of the reconstructed image are recovered. Beyond NIS=3, the improvements diminish, but they accumulate until the reconstructed image is almost equal to the ground-truth image (NIS=50).

\section{Conclusions}
\label{sec:conclu}
We studied the problem of image reconstruction in OAT proposing a reconstruction scheme that makes use of a diffusion model. Starting from an initial reconstructed image using the sinogram, we developed an enhancement image architecture using a conditional diffusion model. In our proposal we used DAS as an initial reconstruction method and a simple autoencoder structure that generates a latent representation of the DAS images to be used as conditional information for the diffussion model for biasing its generative power to the desired image. The scheme is able to provide significant gains with respect to DAS and it is also competitive in terms of image quality with respect to a U-Net architecture. Although the training of this proposal is very demanding in terms of computational resources, once the model is optimized, the computational load required for inference can be easily managed using a small NIS. Future work will be focused in further optimization of the architecture, including the CIP block and the implementation of a full latent diffusion model as the one considered in \cite{Rombach_Blattmann_Lorenz_Esser_Ommer_2022}.


\section*{Acknowledgment}
This work was supported by the University of Buenos Aires (grant UBACYT 20020190100032BA), CONICET (grant PIP 11220200101826CO) and the ANPCyT (grant PICT 2020-01336).


%

\bibliographystyle{IEEEtran}
\bibliography{references}

\end{document}